\documentclass[a4paper,fleqn,usenatbib]{mnras}

\usepackage{newtxtext}
\usepackage{newtxmath}
\usepackage{dblfloatfix}  %

\usepackage[T1]{fontenc}
\usepackage{ae,aecompl}

\usepackage{graphicx}	%

\ifdefined\Bbbk{}
    \let\Bbbk\relax
\else
\fi
\ifdefined\bigtimes{}
    \let\bigtimes\relax
\else
\fi
\usepackage[separate-uncertainty=true,
            multi-part-units=single]{siunitx}
\usepackage{pgf}
\usepackage{xprintlen}
\usepackage{booktabs}

\usepackage{mathabx}  %

\usepackage{wasysym}

\usepackage{sparklines}  %
\usepackage{orcidlink} %
\usepackage{lipsum}  %

\hypersetup{pdfauthor={D. Evensberget et al.}}  %

\newcommand{\batsrus}{{\textsc{bats-r-us}}}
\newcommand{\swmf}{{\textsc{swmf}}}
\newcommand{\awsom}{{\textsc{awsom}}}

\newcommand{\Wind}{\textsc{w}} %
\newcommand{\Rot}{\text{rot}}  %
\newcommand{\Obs}{\text{obs}}  %
\newcommand{\Sat}{\text{sat}}  %
\newcommand{\Alfven}{{\textsc{a}}} %
\newcommand{\Star}{\text{\bigstar}} %

\defcitealias{2013A&A...556A..36G}{GB13}  %

\renewcommand{\Earth}{{\mathchoice{}{}{\scriptscriptstyle}{}\oplus}} %
\renewcommand{\Sun}{{\mathchoice{}{}{\scriptscriptstyle}{}\odot}} %
\renewcommand{\Star}{{\mathchoice{}{}{\scriptscriptstyle}{}\bigstar}} %

\newcommand{\onlineref}[1]{\ref{#1} (on-line supplement)}
\renewcommand{\onlineref}[1]{\ref{#1}}

\DeclareSIUnit\year{yr}
\DeclareSIUnit\astronomicalunit{au}
\DeclareSIUnit\parsec{pc}
\DeclareSIUnit\erg{erg}
\DeclareSIUnit\gauss{G}
\DeclareSIUnit\mSun{\mbox{\(M_\Sun\)}}
\DeclareSIUnit\rSun{\mbox{\(R_\Sun\)}}
\DeclareSIUnit\mEarth{\mbox{\(M_\Earth\)}}
\DeclareSIUnit\rEarth{\mbox{\(R_\Earth\)}}

\renewcommand{\vec}[1]{\boldsymbol{#1}} %
\newcommand{\uvec}[1]{\boldsymbol{\hat{#1}}} %

\newcommand\thefont{\expandafter\string\the\font} %

\renewcommand{\propto}{\mathrel{\mathchar"939}} %

\newcommand{\appropto}{\mathrel{\vcenter{  %
  \offinterlineskip\halign{\hfil\(##\)\cr   %
    \propto\cr\noalign{\kern2pt}\sim\cr\noalign{\kern-2pt}}}}}   %

\usepackage{star-commands}

\title[Rotational evolution with data-driven 3D winds]
    {Rotational evolution of young-to-old stars with data-driven three-dimensional wind models}
\author[D. Evensberget et al.]{
    D. Evensberget \orcidlink{0000-0001-7810-8028}\(^{1}\)\thanks{E-mail: \href{mailto:evensberget@strw.leidenuniv.nl}{evensberget@strw.leidenuniv.nl}},
    A. A. Vidotto  \orcidlink{0000-0001-5371-2675}\(^{1}\),
    \\
    \(^{1}\)Leiden Observatory, Leiden University, PO Box 9513, 2300 RA Leiden, The Netherlands
}

\date{Accepted XXX.\@ Received YYY;\@ in original form ZZZ}
\pubyear{2024}

\begin{document}
\label{firstpage}  %
\pagerange{\pageref{firstpage}--\pageref{lastpage}}
\maketitle
\begin{abstract}
    Solar-type stars form with a wide range of rotation rates $\Omega$. A wide $\Omega$ range persists until a stellar age of $t\sim 0.6$ Gyr, after which solar-type stars exhibit Skumanich spin-down where $\Omega \propto t^{-1/2}$.
Rotational evolution models incorporating polytropic stellar winds struggle to simultaneously reproduce these two regimes, namely the initially wide $\Omega$ range and the Skumanich spin-down without imposing an a-priori cap on the wind mass-loss rate.
We show that a three-dimensional wind model driven by Alfvén waves and observational data yields wind torques that agree with the observed age distribution of $\Omega$.
In our models of the Sun and twenty-seven open cluster stars aged from 0.04 to 0.6 Gyr that have observationally derived surface magnetic maps and rotation rates, we find evidence of exponential spin-down in young stars that are rapid rotators and Skumanich spin-down for slow rotators. The two spin-down regimes emerge naturally from our data-driven models. Our modelling suggests that the observed age distribution of stellar rotation rates $\Omega$ arises as a consequence of magnetic field strength saturation in rapid rotators.

\end{abstract}


\section{Introduction}
Solar-type (FGK) stars enter the stellar spin-down era with a wide range of rotation rates
\(\Omega\)~\citep{1985ApJ...289..247S,1991ASIC..340...41B,1993AJ....106..372E}.
The fastest rotators spin near the stellar break-up velocity and the slowest rotators are up to two orders of magnitude slower~\citep{2013A&A...556A..36G}.
Observations of stellar clusters  show that this wide range of periods is eventually lost; some mechanism harmonises their rotation rates by the onset of the spin-down era at an age \(t_\textsc{s}\sim\qty{0.6}{\giga\year}\) ~\citep{1972ApJ...171..565S}, after which \(\Omega \propto t^{-1/2}\).
To reproduce the observed \(\Omega\) distribution, spin-down models have incorporated a `knee' in the \(\dot \Omega(\Omega)\) relation~\citep[e.g.\ ][]{1987ApJ...318..337S,1995ApJ...441..865C,1997ApJ...480..303K}, so that \(\dot \Omega\propto \Omega\) for rapid rotators and \(\dot \Omega\propto \Omega^{3}\) for slow rotators as required to attain Skumanich spin-down~\citep[e.g.\ ][]{1978GApFD...9..241D}.

Solar-type stars spin-down by means of magnetised stellar winds~\citep[e.g.\ ][]{1958ApJ...128..664P,1962AnAp...25...18S} both before and during the Skumanich spin-down era. From one-dimensional models, we can write the wind torque as \(\dot J \propto \dot M \Omega R_\Alfven^n\) where \(\dot M\) is the wind mass-loss rate and \(R_\Alfven\) is the Alfvén radius~\citep{1967ApJ...148..217W,1968MNRAS.140..177M,1984LNP...193...49M,1988ApJ...333..236K}.
Mass-loss rate saturation in rapid rotators~\citep[e.g.\ ][]{2011ApJ...741...54C} can be imposed to produce the required knee in the \(\dot \Omega(\Omega)\) relation. In polytropic magnetohydrodynamic wind models~\citep{2009ApJ...699..441V,2012ApJ...754L..26M,2015ApJ...798..116R,2017ApJ...845...46F,2018ApJ...864..125F}, the wind mass loss rate is governed largely by the coronal temperature and wind density. By varying the wind density, and thus \(\dot M\), models are able to match spin-down `gyrotracks' \(\Omega(t)\) found by integrating \(\dot \Omega\) to observational data~\citep{2013A&A...556A..36G,2015A&A...577A..28J,2017MNRAS.472.2590S}. This mass loss rate saturation is, however, externally imposed on the wind model. In contrast to polytropic wind models, an Alfvén wave based wind model~\citep[][]{2013ApJ...764...23S,2014ApJ...782...81V} can shift the wind model boundary downwards into the much colder and denser stellar chromosphere, and compute the coronal wind density and \(\dot M\) consistently with \(\dot J\) and \(R_\Alfven\) \citep[e.g.\ ][]{2015ApJ...807L...6G,2016A&A...588A..28A,2017ApJ...835..220C}.

In this letter, we show that the resulting \(\dot \Omega\) of the Sun and a set of twenty-seven young, solar-type cluster stars aged from \qtyrange{\sim 40}{\sim600}{\mega\year}~\citep{paper1,paper2,paper3} appears compatible with the observed age distribution of stellar rotation rates \(\Omega\)~\citep{2013A&A...556A..36G} for a solar mass star without imposing any {a priori} assumptions about mass loss saturation.
In Section~\ref{sec:analysis}, we show this by comparing instantaneous wind torque \(\dot J_\Wind\) of our models to a simple observation-based gyrotrack model based on the work of \citet{2013A&A...556A..36G} containing hundreds of rotation rates.
The resulting wind torques do however need to be scaled by a factor of \(\sim 7\) in order to yield the expected Skumanich spin-down rate at solar maximum conditions, as presented in Section~\ref{sec:Discussion}. This scaling factor is the only free parameter in our calculated wind torques.
In Section~\ref{sec:Conclusions} we present our conclusions.

\section{Wind torque model}\label{sec:analysis}

In our recent work we have modelled the winds of thirty
young, solar-type stars~\citep{paper1,paper2,paper3} with well characterised ages, and whose  surface magnetic fields
have previously been mapped with Zeeman-Doppler imaging (ZDI) by \citet{2016MNRAS.457..580F,2018MNRAS.474.4956F}. The modelled stars are part of the Hyades, Coma Berenices, Hercules-Lyra, Pleiades, AB Doradus, and Columba-Horologium-Tucana clusters\footnote{
    The \(\beta\) Pictoris stars studied in~\citet{paper3} are excluded from the analysis in this work as their age of \qty{24\pm 3}{\mega\year} suggests that they are still contracting and have not yet reached the spin-down era.
}. As cluster members, these stars' ages are known to within \qtyrange{5}{20}{\percent}. For reference the stars' ages, rotation rates, masses and radii are given in Table~\onlineref{tab:observed_quantities}. This is the largest sample of stars with well characterised ages and ZDI surface magnetic fields derived with a consistent methodology.

For each star\label{text:numerical model} a wind model was obtained by letting the surface magnetic field
drive a three-dimensional numerical wind model~\citep{1999JCoPh.154..284P, 2012JCoPh.231..870T} incorporating Alfvén wave coronal heating~\citep{2013ApJ...764...23S,2014ApJ...782...81V}. The resulting wind solutions permit us to simultaneously compute consistent key quantities such as
wind mass loss rate \(\dot M_\Wind\) and
wind angular momentum loss rate \(\dot J_\Wind\) as described in Appendix~\onlineref{sec:wind-models}.
Identical solar parameter values are used in all the wind models; the only varying parameters are the surface magnetic map, stellar rotation rate, mass, and radius as indicated in Table~\onlineref{tab:parameters}.

\begin{figure*}
    \includegraphics[scale=1.5]{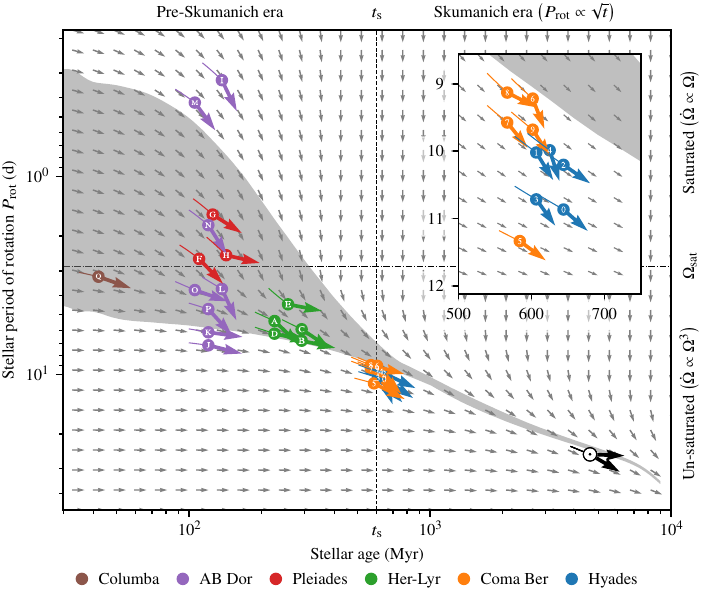}
    \caption{
        Stellar rotational period and wind-derived spin-down rates \(\dot \Omega_\Wind\)
        plotted against stellar age for the stellar wind models in \citet{paper1,paper2,paper3}.
        The angular velocity of each star is plotted against the star's age with a coloured symbol. The spin-down rate \(\dot \Omega_\Wind\) is indicated with an arrow extending from the symbol. Larger values of \(\dot J_\Wind\) and thus of \(\dot \Omega_\Wind\) produce more downwards-pointing arrows in the plot.
        In order for the symbols not to overlap each other we have added a small pseudorandom shift to the stellar ages.
        As the Hyades %
        and Coma Berenices %
        stars are close together in both \(\Omega\) and age, we have added an inset in the top right corner which gives a zoomed in view of this region.
        The Sun symbol and the two black arrows extending from it represent the solar minimum and solar maximum models from \citet{paper3}.
        In the plot background we have added a shaded region corresponding to the region between the 25\textsuperscript{th} and 90\textsuperscript{th} percentile of stellar \(\Omega_\Obs(t)\) gyrotracks as a function of age for a solar mass star from \citet{2013A&A...556A..36G}.
        The plot background also contains a vector field (grey arrows) of \(\dot\Omega_\Obs\) values from the fitted model described by equation~\eqref{eq:omegadot-from-obs}.
        The vertical dashed line denotes the onset of the Skumanich spin-down era, while the horizontal dash-dotted line separates the saturated and un-saturated spin-down regimes as noted in the top and right margins.
        }\label{fig:trend-omega-age}
\end{figure*}

\begin{table*}
    \centering
    \sisetup{
        table-figures-decimal=2,
        table-figures-integer=2,
        table-figures-uncertainty=2,
        table-figures-exponent = 0,
        table-number-alignment=center,
        round-mode=places,
        round-precision=1
    }
    \caption{
        Relevant model parameters and results from the wind models~\citep{paper3} used in this work and stellar rotational inertia values.
        The `case' column denotes the model case number from~\citet{paper3}. The `type' column gives the star's spectral type. When the spectral type is calculated from \(T_\text{eff}\) following~\citet{2013ApJS..208....9P} this is denoted by a dagger symbol (\(\dag\)).
        The `age' column gives the stellar age, and the following columns give the stellar mass \(M\), radius \(R\), and rotational period \(P_\text{rot}\). The \(B_r\) value is the average unsigned radial magnetic field over the stellar surface from the Zeeman-Doppler maps of~\citet{2016MNRAS.457..580F,2018MNRAS.474.4956F}, from which the the previous columns values were also taken.
        The wind mass loss rate \(\dot M_\Wind\) and wind torque \(\dot J_\Wind\) are computed from the wind models. A scaling factor of 7 has been applied to the \(\dot J_\Wind\) values as described in Section~\ref{sec:torque-scaling}. In the main text this factor is incorporated into the definition of \(\dot J_\Wind\).
        The rotational inertia \(I\) values are taken from the models of~\citet{2015A&A...577A..42B}. The stellar age column can also be used to identify the parent cluster as follows:
        \sisetup{round-precision=0}
        Columba-Horologium-Tucana:  \qty{42\pm 6}{\mega\year},
        AB Doradus:                 \qty{120\pm 10}{\mega\year},
        Pleiades:                   \qty{125\pm 8}{\mega\year},
        Hercules-Lyra:              \qty{257\pm 46}{\mega\year},
        Coma Berenices:             \qty{584\pm 10}{\mega\year},
        Hyades:                     \qty{625\pm 50}{\mega\year}.
    }\label{tab:observed_quantities}
    \begin{tabular}{
    l
    l
    S[table-format=3(2), round-precision=0]  %
    S[table-format=1.2, round-precision=2]  %
    S[table-format=1.2, round-precision=2]  %
    S[table-format=2.2, round-precision=2]  %
    S[table-format=2.1, round-precision=1]  %
    S[table-format=2.1, round-precision=1]  %
    S[table-format=1.2e2, round-precision=2]  %
    S[table-format=1.2e2, round-precision=2]  %
    S[table-format=1.2e2, round-precision=2]  %
}

\toprule
Case
& {Type}
& {Age}
& {\(M\)}
& {\(R\)}
& {\(P_\Rot\)}
& {\(B_r\)}
& {\(R_\Alfven\) }
& {\(\dot M_\Wind\) }
& {\(7\times \dot J_\Wind\) }
& {\(I\)}
\\ %
&
& {\( \left(  \si{\mega\year}                 \right) \)}
& {\( \left(  M_\Sun                    \right) \)}
& {\( \left(  R_\Sun                \right) \)}
& {\( \left(  \si{\day}               \right) \)}
& {\( \left(  \si{\gauss}                \right) \)}
& {\( \left(  R_\Star{}                  \right) \)}
& {\( \left(  \si{\kilogram\per\second}  \right) \)}
& {\( \left(  \si{\newton\meter}         \right) \)}
& {\( \left(  \si{\kilogram\meter\squared}               \right) \)}
\\
\midrule
\SimSymbol{1}{0} \href{http://simbad.u-strasbg.fr/simbad/sim-id?Ident=Cl+Melotte+25+5         }{Cl Melotte 25 5         }                  & {K0}       &  625\pm 50 & 0.85       & 0.91       & 10.57      & 5.96       & 12.0       & 4.95e+09   & 7.80e+24   & 3.66e+46  \\
\SimSymbol{1}{1} \href{http://simbad.u-strasbg.fr/simbad/sim-id?Ident=Cl+Melotte+25+21        }{Cl Melotte 25 21        }                  & {G5}       &  625\pm 50 & 0.90       & 0.91       & 9.73       & 9.97       & 14.4       & 7.54e+09   & 1.68e+25   & 4.27e+46  \\
\SimSymbol{1}{2} \href{http://simbad.u-strasbg.fr/simbad/sim-id?Ident=Cl+Melotte+25+43        }{Cl Melotte 25 43        }                  & {K0}       &  625\pm 50 & 0.85       & 0.79       & 9.90       & 5.80       & 12.4       & 3.31e+09   & 4.68e+24   & 2.79e+46  \\
\SimSymbol{1}{3} \href{http://simbad.u-strasbg.fr/simbad/sim-id?Ident=Cl+Melotte+25+151       }{Cl Melotte 25 151       }                  & {K2}       &  625\pm 50 & 0.85       & 0.82       & 10.41      & 10.87      & 13.8       & 6.61e+09   & 9.92e+24   & 2.97e+46  \\
\SimSymbol{1}{4} \href{http://simbad.u-strasbg.fr/simbad/sim-id?Ident=Cl+Melotte+25+179       }{Cl Melotte 25 179       }                  & {K0}       &  625\pm 50 & 0.85       & 0.84       & 9.70       & 17.03      & 16.2       & 1.05e+10   & 2.35e+25   & 3.11e+46  \\
\midrule
\SimSymbol{1}{5} \href{http://simbad.u-strasbg.fr/simbad/sim-id?Ident=TYC+1987-509-1          }{TYC 1987-509-1          }                  & {K2}       &  584\pm 10 & 0.80       & 0.72       & 11.10      & 10.06      & 12.9       & 4.10e+09   & 3.47e+24   & 1.94e+46  \\
\SimSymbol{1}{6} \href{http://simbad.u-strasbg.fr/simbad/sim-id?Ident=Cl*+Melotte+111+AV+1693 }{Cl* Melotte 111 AV 1693 }                  & {G8}       &  584\pm 10 & 0.90       & 0.83       & 9.05       & 20.34      & 14.8       & 1.15e+10   & 2.25e+25   & 3.62e+46  \\
\SimSymbol{1}{7} \href{http://simbad.u-strasbg.fr/simbad/sim-id?Ident=Cl*+Melotte+111+AV+1826 }{Cl* Melotte 111 AV 1826 }                  & {G9}       &  584\pm 10 & 0.85       & 0.80       & 9.34       & 12.57      & 12.7       & 7.33e+09   & 8.50e+24   & 2.84e+46  \\
\SimSymbol{1}{8} \href{http://simbad.u-strasbg.fr/simbad/sim-id?Ident=Cl*+Melotte+111+AV+2177 }{Cl* Melotte 111 AV 2177 }                  & {G6}       &  584\pm 10 & 0.90       & 0.78       & 8.98       & 5.48       & 11.7       & 2.76e+09   & 3.86e+24   & 3.13e+46  \\
\SimSymbol{1}{9} \href{http://simbad.u-strasbg.fr/simbad/sim-id?Ident=Cl*+Melotte+111+AV+523  }{Cl* Melotte 111 AV 523  }                  & {G7}       &  584\pm 10 & 0.90       & 0.83       & 9.43       & 14.61      & 14.0       & 8.27e+09   & 1.33e+25   & 3.56e+46  \\
\midrule
\SimSymbol{1}{A} \href{http://simbad.u-strasbg.fr/simbad/sim-id?Ident=DX+Leo                  }{DX Leo                  }                  & {G9}       &  257\pm 46 & 0.90       & 0.81       & 5.38       & 21.05      & 17.8       & 1.14e+10   & 5.65e+25   & 3.42e+46  \\
\SimSymbol{1}{B} \href{http://simbad.u-strasbg.fr/simbad/sim-id?Ident=EP+Eri                  }{EP Eri                  }                  & {K1}       &  257\pm 46 & 0.85       & 0.72       & 6.76       & 9.48       & 11.7       & 3.97e+09   & 5.85e+24   & 2.28e+46  \\
\SimSymbol{1}{C} \href{http://simbad.u-strasbg.fr/simbad/sim-id?Ident=HH+Leo                  }{HH Leo                  }                  & {G8}       &  257\pm 46 & 0.95       & 0.84       & 5.92       & 15.29      & 15.8       & 9.32e+09   & 3.65e+25   & 4.26e+46  \\
\SimSymbol{1}{D} \href{http://simbad.u-strasbg.fr/simbad/sim-id?Ident=V439+And                }{V439 And                }                  & {K0}       &  257\pm 46 & 0.95       & 0.92       & 6.23       & 8.74       & 13.4       & 6.85e+09   & 1.87e+25   & 5.05e+46  \\
\SimSymbol{1}{E} \href{http://simbad.u-strasbg.fr/simbad/sim-id?Ident=V447+Lac                }{V447 Lac                }                  & {K1}       &  257\pm 46 & 0.90       & 0.81       & 4.43       & 10.40      & 12.6       & 5.84e+09   & 1.40e+25   & 3.42e+46  \\
\midrule
\SimSymbol{1}{F} \href{http://simbad.u-strasbg.fr/simbad/sim-id?Ident=HII+296                 }{HII 296                 }                  & {G8}       &  125\pm  8 & 0.90       & 0.94       & 2.61       & 51.40      & 22.5       & 3.14e+10   & 3.42e+26   & 4.50e+46  \\
\SimSymbol{1}{G} \href{http://simbad.u-strasbg.fr/simbad/sim-id?Ident=HII+739                 }{HII 739                 }                  & {G0}       &  125\pm  8 & 1.15       & 1.07       & 1.56       & 49.60      & 16.3       & 3.67e+10   & 8.20e+26   & 1.13e+47  \\
\SimSymbol{1}{H} \href{http://simbad.u-strasbg.fr/simbad/sim-id?Ident=Cl*+Melotte+22+P        }{Cl* Melotte 22 PELS 31  }                  & {K2\textsuperscript{\dag}} &  125\pm  8 & 0.95       & 1.05       & 2.50       & 17.03      & 11.8       & 1.71e+10   & 1.01e+26   & 6.58e+46  \\
\midrule
\SimSymbol{1}{I} \href{http://simbad.u-strasbg.fr/simbad/sim-id?Ident=BD-07+2388              }{BD-07 2388              }                  & {K0}       &  120\pm 10 & 0.85       & 0.78       & 0.33       & 94.16      & 30.2       & 3.54e+10   & 2.23e+27   & 2.66e+46  \\
\SimSymbol{1}{J} \href{http://simbad.u-strasbg.fr/simbad/sim-id?Ident=HD+6569                 }{HD 6569                 }                  & {K1}       &  120\pm 10 & 0.85       & 0.76       & 7.13       & 15.27      & 16.1       & 7.36e+09   & 1.71e+25   & 2.50e+46  \\
\SimSymbol{1}{K} \href{http://simbad.u-strasbg.fr/simbad/sim-id?Ident=HIP+10272               }{HIP 10272               }                  & {K1}       &  120\pm 10 & 0.90       & 0.80       & 6.13       & 10.07      & 14.5       & 5.11e+09   & 1.44e+25   & 3.29e+46  \\
\SimSymbol{1}{L} \href{http://simbad.u-strasbg.fr/simbad/sim-id?Ident=HIP+76768               }{HIP 76768               }                  & {K5}       &  120\pm 10 & 0.80       & 0.85       & 3.70       & 58.64      & 23.1       & 3.05e+10   & 2.02e+26   & 2.66e+46  \\
\SimSymbol{1}{M} \href{http://simbad.u-strasbg.fr/simbad/sim-id?Ident=LO+Peg                  }{LO Peg                  }     (HIP 106231) & {K3}       &  120\pm 10 & 0.75       & 0.66       & 0.42       & 88.58      & 29.0       & 2.19e+10   & 8.13e+26   & 1.34e+46  \\
\SimSymbol{1}{N} \href{http://simbad.u-strasbg.fr/simbad/sim-id?Ident=PW+And                  }{PW And                  }                  & {K2}       &  120\pm 10 & 0.85       & 0.78       & 1.76       & 70.67      & 24.7       & 2.49e+10   & 3.90e+26   & 2.68e+46  \\
\SimSymbol{1}{O} \href{http://simbad.u-strasbg.fr/simbad/sim-id?Ident=TYC+0486-4943-1         }{TYC 0486-4943-1         }                  & {K4\textsuperscript{\dag}} &  120\pm 10 & 0.77       & 0.70       & 3.75       & 18.72      & 14.9       & 8.61e+09   & 3.25e+25   & 1.61e+46  \\
\SimSymbol{1}{P} \href{http://simbad.u-strasbg.fr/simbad/sim-id?Ident=TYC+5164-567-1          }{TYC 5164-567-1          }                  & {K1\textsuperscript{\dag}} &  120\pm 10 & 0.90       & 0.90       & 4.68       & 48.50      & 21.6       & 2.96e+10   & 1.68e+26   & 4.13e+46  \\
\midrule
\SimSymbol{1}{Q} \href{http://simbad.u-strasbg.fr/simbad/sim-id?Ident=BD-16+351               }{BD-16 351               }                  & {K1\textsuperscript{\dag}} &   42\pm  6 & 0.90       & 0.96       & 3.21       & 32.77      & 20.5       & 2.37e+10   & 3.08e+26   & 5.07e+46  \\
\midrule
{\(\odot\)} Sun CR2157 (solar maximum) & {G2}       & 4600\pm  0 & 1.00       & 1.00       & 25.38      & 5.55       & 6.9        & 3.84e+09   & 7.95e+23   & 7.06e+46  \\
{\(\odot\)} Sun CR2211 (solar minimum) & {G2}       & 4600\pm  0 & 1.00       & 1.00       & 25.38      & 0.61       & 5.0        & 3.94e+08   & 5.55e+22   & 7.06e+46  \\
\bottomrule
\end{tabular}

\end{table*}
\begin{table}
	\centering
	\caption{
        Wind modelling parameters used for all the stellar and solar models used in this work. The parameters that vary between wind models are the surface magnetic map, stellar rotation rate, mass, and radius. All other modelling parameters are fixed. Note that the stellar age is not a wind model input.
    }\label{tab:parameters}
	\begin{tabular}{ll}
		\toprule
		Parameter & Value \\
       \midrule
       Stellar surface magnetic map & See \citet{2016MNRAS.457..580F,2018MNRAS.474.4956F} \\
       Stellar rotation rate        & Table~\ref{tab:observed_quantities} \\
       Stellar mass                 & Table~\ref{tab:observed_quantities} \\
       Stellar radius               & Table~\ref{tab:observed_quantities} \\
       \midrule
       Chromospheric base temperature & \qty{5.0e4}{\kelvin} \\
       Chromospheric base density & \qty{2.0e17}{\per\cubic\meter}  \\
       Poynting flux-to-field ratio & \qty{1.1e6}{\watt\per\square\meter\per\tesla}  \\
       Turbulence correlation length & \qty{1.5e5}{\meter\tesla^{1/2}} \\
	   \bottomrule
	\end{tabular}
\end{table}

\subsection{Rotational braking from wind torques}\label{sec:wind-torques}
The stellar spin-down rates \(\dot \Omega_\Wind\) are computed from (scaled) wind torques \(\dot J_\Wind\) that we previously published~\citep{paper3}. In this paper we apply a scaling factor of 7 to our stellar and solar \(\dot J_\Wind\) values to match the expected Skumanich spin-down rate at solar maximum conditions.

To compute spin-down rates from wind torques, we use the stellar rotational inertia values \(I\) from~\citet{2015A&A...577A..42B}. As \(J = \Omega I\) we get~\citep[e.g.\ ][]{2015A&A...577A..28J,2015ApJ...799L..23M}
\begin{equation}\label{eq:omegadot-from-torque}
    \dot \Omega_\Wind = \left. \dot J_\Wind \middle/ I - \Omega \, \dot I \middle/ I\right.
    \qquad \text{(Stellar wind models)}
\end{equation}
where \(\dot J_\Wind\) is the wind torque and \(\Omega \, \dot I\) is the contraction torque~\citep{2015A&A...577A..98G}. In the \citeauthor{2015A&A...577A..42B} models \(I\) is essentially steady by \(t \sim \qty{100}{\mega\year}\) so that the wind torque term dominates \(\dot \Omega_\Wind\). For convenience, the \(I\) and values used are given in Appendix~\onlineref{sec:spin-down-timescales}.
In Fig.~\ref{fig:trend-omega-age} the stellar rate of rotation is plotted against the stellar age (coloured circles). The stellar spin-down rates \(\dot \Omega_\Wind\) derived from our wind models are shown as coloured arrow attached to the symbol of each star, and pointing in the direction of \(\dot \Omega_\Wind\).

To guide the eye, the shaded region of Fig.~\ref{fig:trend-omega-age} shows the range between the 25\textsuperscript{th} and 90\textsuperscript{th} percentile of observed \(\Omega\) values~\citep[curves from][]{2013A&A...556A..36G}.
The pre-Skumanich spin-down era may be identified as the time where there is significant difference between the two percentiles (up to \(t\sim \qty{600}{\mega\year}\)). This era is followed by the Skumanich spin-down era where \(\Omega \propto t^{-1/2}\).
From Fig.~\ref{fig:trend-omega-age}
we have good coverage of the \citet{2013A&A...556A..36G} sample around \qty{0.1}{\giga\year} and \qty{0.6}{\giga\year}, however our ZDI dataset has no stars between onset of the Skumanich relation and the solar models.

\subsection{Rotational braking from cluster spin rates}\label{sec:gyrotracks}
\citet{2013A&A...556A..36G} fitted \(\Omega(t)\) curves, so-called `gyrotracks', to the observed rotation rates of young and very young (\(t < \qty{30}{\mega\year}\)) clusters, incorporating mass loss saturation, core-envelope decoupling, and other effects.
recently \emph{GAIA} data has been used to infer the age distribution of rotation rates for a much larger sample of stars. The essential features of the revised gyrotracks are however retained~\citep[e.g.][Fig.~13]{2021ApJS..257...46G} in the age and mass range where we have wind model data. Here, we fit a very simple model on the form
\begin{equation}\label{eq:omegadot-from-obs}
    \dot \Omega_\Obs =
    \left\{
        \begin{aligned}
            -k_1 \Omega\phantom{^3}&, \quad \Omega >    \Omega_\Sat  \\
            -k_2 \Omega        {^3}&, \quad \Omega \leq \Omega_\Sat  \\
        \end{aligned}, \qquad \text{(Vector field)}
    \right.
\end{equation}
to the observation-based \citet{2013A&A...556A..36G} 25\textsuperscript{th} and 90\textsuperscript{th} percentile \(\Omega(t)\) gyrotrack curves, focusing on the spin-down era
\(
\qty{30}{\mega\year}
<
t
<
\qty{4.6}{\giga\year}
\)
for a one solar mass star\footnote{
    The grey band in Fig.~\ref{fig:trend-omega-age} is for a 1 \(M_\Sun\) star, while many of the stars modelled in this study mass around \(\qty{0.85\pm0.10}\, M_\Sun\). We do not expect this to affect the qualitative conclusions in this paper since the gyrotracks are qualitatively similar for a \(0.8 M_\Sun\) star in \citet{2015A&A...577A..98G}.
}.
We note that much of the signal for solar analogues is in stars less than \qty{100}{\mega\year} old \citep{2010ApJ...716.1269D,2013A&A...556A..36G,2015A&A...577A..98G}.

Our simplified model~\eqref{eq:omegadot-from-obs} permits gyrotracks \(\Omega_\Obs(t)\) to decay exponentially when \(\Omega > \Omega_\Sat\), while adhering to the \citet{1972ApJ...171..565S} relation \(\Omega(t) \propto t^{-1/2}\) for \(\Omega < \Omega_\Sat\). We have parametrised the pre-Skumanich spin-down era in terms of an exponential decay constant \(k_1\) so that the rotational spin-down timescale in this era \(\Omega / \dot \Omega_\Obs = k_1^{-1} = \qty{120}{\mega\year}\).
The value of \(k_2\) is then found by matching the solar age and angular velocity:
\(k_2 = \frac12 \Omega_\Sun^{-2} t_\Sun^{-1} = \qty{4.1e-7}{\second\per\square\radian}\). From continuity of \(\dot \Omega_\Obs\) the saturation threshold is \(\Omega_\Sat = \sqrt{k_1/k_2}\). This gives a reasonable \(\Omega_\Sat = 8.9\,\Omega_\Sun\), given that this parameter can vary between \(3\,\Omega_\Sun\) and \(15\,\Omega_\Sun\) depending on model choices~\citep{2015ApJ...799L..23M,2016A&A...587A.105A}. In Fig.~\ref{fig:trend-omega-age} the resulting \(\dot \Omega_\Obs\) values are shown as a vector field (little grey arrows).
By design we have excellent agreement between the 25\textsuperscript{th} and 90\textsuperscript{th} percentile \(\Omega(t)\) gyrotrack curves (edges of the grey region) and the \(\dot \Omega_\Obs\) vector field (grey arrows), indicating that eq.~\eqref{eq:omegadot-from-obs} captures the dominant features of the \citet{2013A&A...556A..36G} spin-down era gyrotracks for a Solar-mass star.
In addition to the vector field, we have also added small trailing gyrotrack segments \(\Omega_\Obs(t)\) to each star. By comparing the slope of each stars' trailing trajectory segment \(\Omega_\Obs(t)\) to its emerging arrow \(\dot \Omega_\Wind\) it can be seen that there is good agreement between the wind torque based spin-down rates and the gyrotrack segments.

As mentioned in Section~\ref{sec:wind-torques}, we apply a scaling factor of \qty{7} to the wind torques \(\dot J_\Wind\) of \citet{paper3}. In our dataset this corresponds to matching the wind torque \(\dot J_\Wind\) at solar maximum to the Sun's gyrotrack predicted torque \(\dot J_\Obs\). The scaling effect can be seen in Fig.~\ref{fig:trend-omega-age} where the most downwards-pointing of the wind torque arrows extending from the Sun symbols is in good agreement with the gyrotrack trail. The wind torque scaling is discussed further in Section~\ref{sec:Discussion}.

\subsection{Saturation in rapid rotators}
\begin{figure}
    \centering
    \includegraphics{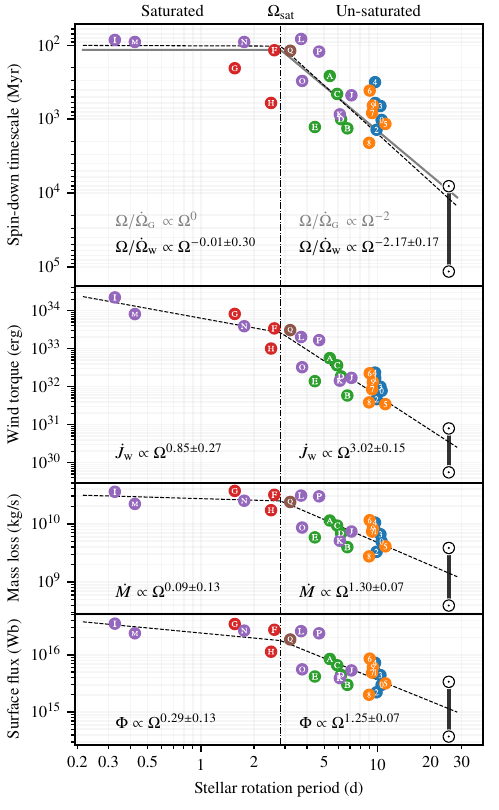}
    \caption{
        In the top panel we compare the spin-down timescale inferred from the wind models \(\Omega / \dot \Omega_\Wind\) (coloured symbols) and the spin-down timescale \(\Omega / \dot \Omega_\Obs\) (grey curve) from equation~\eqref{eq:omegadot-from-obs}. The two Sun symbols connected by a black bar represents the solar maximum and minimum.
        The second panel shows the wind torque \(\dot J_\Wind\) (\(\qty{1}{erg}=\qty{e-7}{\newton\meter}\)) and
        the third panel shows the wind mass loss rate \(\dot M_\Wind\) values (the commonly used solar mass loss rate value is
        \(\num{2e-14} M_\Sun \, \si{\per\year} = \SI{1e9}{\kilogram\per\second}\)).
        The bottom panels show the surface unsigned magnetic flux \(\Phi\) which is determined directly from the magnetic maps driving the numerical models.
        The dashed black lines show the broken power law fit of equation~\eqref{eq:broken-power-fit}.
        The vertical dash-dotted line shows \(\Omega_\Sat = 8.9 \, \Omega_\Sun\) which separates the saturated (\(\Omega > \Omega_\Sat\)) and un-saturated (\(\Omega < \Omega_\Sat\)) spin-down regimes.
    }\label{fig:detailed-breakdown}
\end{figure}

The good agreement between  the wind torque based spin-down rates \(\dot \Omega_\Wind\) (coloured arrows) and the gyrotracks \(\dot \Omega_\Obs(t)\) (coloured trailing curve segments) in Fig.~\ref{fig:trend-omega-age} suggests the presence of a saturation effect in the \(\dot \Omega_\Wind\) values around \(\Omega_\Sat\). This is confirmed in Fig.~\ref{fig:detailed-breakdown} where
the top panel shows the calculated spin-down timescale \(\Omega / \dot \Omega_\Wind\) plotted against the rotation period as coloured symbols, and \(\Omega / \dot \Omega_\Obs\) (see equation~\ref{eq:omegadot-from-obs}) as a grey curve  consisting of two line segments. Vertical distance between the star symbols and the gray curve correspond to trail-arrow mismatch in Fig.~\ref{fig:trend-omega-age}.

The dashed black curve in each panel shows a fitted broken power law
\begin{equation}\label{eq:broken-power-fit}
    y / y_\Sat = \left(\Omega / \Omega_\Sat\right)^{a}
\end{equation}
in which \(\Omega_\Sat = 8.9 \,\Omega_\Sun\) is fixed, while \(y_\Sat\) and \(a\) are found by simultaneous fitting. The exponent \(a\) is permitted to take on a different value on each side of \(\Omega_\Sat\). The \(a\) and \(y_\Sat\) values are given in Table~\onlineref{tab:fit-coefficients}, and the \(a\) values are annotated in Fig.~\ref{fig:detailed-breakdown}.
\begin{table}
    \caption{Fitted coefficients for the broken power law \(y / y_\Sat = \left(\Omega / \Omega_\Sat\right)^{a}\). The first two numerical columns give the exponent \(a\) values on either side of the saturation point \(\Omega_\Sat = 8.9\, \Omega_\Sun\). The final column gives the value of \(\log_{10} y_\Sat\) for \(y_\Sat = y(\Omega_\Sat)\).
    The first three of these relationships are plotted in Fig.~\ref{fig:detailed-breakdown}.
    The average Alfvén radius \(R_\Alfven\) is plotted in Fig.~\ref{fig:detailed-breakdown-alfven}.
    The coefficient of determination \(r^2\) is given in the final column.
     }\label{tab:fit-coefficients}
    \centering
    \begin{tabular}{
        l
        S[table-format=1.2(2),bracket-ambiguous-numbers=false]
        S[table-format=1.2(2),bracket-ambiguous-numbers=false]
        S[table-format=1.2(2),bracket-ambiguous-numbers=false]
        S[table-format=1.2]
    }
    \toprule
    Quantity
    & {\(a\) (\(\Omega > \Omega_\Sat\))}
    & {\(a\) (\(\Omega < \Omega_\Sat\))}
    & {\(\log_{10} y_\Sat\)}
    & {\(r^2\)}
    \\
    \midrule
    {\(\Omega / \dot \Omega_\textsc{w}  \)} ({$\unit{\mega\year}$})&  -0.01 \pm  0.30& -2.17 \pm  0.17&  2.01 \pm  0.04&  0.72\\  %
{\(\dot J_\textsc{w}                \)} ({$\unit{\newton\meter}$})&   0.85 \pm  0.27&  3.02 \pm  0.15& 26.42 \pm  0.04&  0.87\\  %
{\(\dot M                           \)} ({$\unit{\kilogram\per\second}$})&   0.09 \pm  0.13&  1.30 \pm  0.07& 10.39 \pm  0.02&  0.70\\  %
{\(\Phi                             \)} ({$\unit{\weber}$})&   0.29 \pm  0.13&  1.25 \pm  0.07& 16.25 \pm  0.02&  0.70\\  %
{\(B_r                              \)} ({$\unit{\gauss}$})&   0.39 \pm  0.12&  1.31 \pm  0.07&  1.62 \pm  0.02&  0.74\\  %
{\(R_\textsc{a}                     \)} ({$R$})   &   0.14 \pm  0.02&  0.49 \pm  0.01&  1.33 \pm  0.00&  0.75    %
  \\  %
    \bottomrule
    \end{tabular}
\end{table}
From its top panel it can be seen that the fitted curve closely matches the observed \(\Omega/\dot\Omega_\Obs\) relation of equation~\eqref{eq:omegadot-from-obs} shown and annotated in grey.

The middle panel of Fig.~\ref{fig:detailed-breakdown} show the wind mass loss rate \(\dot M_\Wind\) plotted against the rotation period.
The wind mass loss rate exhibits a similar pattern of saturation (notwithstanding a slightly different scatter) around \(\Omega_\Sat\) as that of the spin-down timescale. The mass loss rate \(\dot M\) appears to saturate for \(\Omega>\Omega_\Sat\) in a similar manner as \(\Omega / \dot \Omega_\Wind\).

The bottom panel of Fig.~\ref{fig:detailed-breakdown} shows the surface unsigned magnetic flux
\(\Phi = 4\pi R^2 \, B_r\)
plotted against the rotation period\footnote{By \(B_r\) we mean the average strength of the surface radial magnetic field.}. Once again the fitted dashed curve and the scatter have a similar shape those of the preceding panels. Hence both the \(\dot M_\Wind\) and \(\Phi\) values exhibit similar trends as the top panel \(\Omega/\dot \Omega_\Wind\) values, albeit with different slopes and intersections with the range of solar values.

\section{Discussion}\label{sec:Discussion}

Multiple, possibly interacting physical phenomena have been proposed as causes for the torque saturation in rapid rotators. These include a decoupling mechanism where the stellar convective zone may rotate at a different rate than the radiative zone~\citep{1986PASP...98.1233S,1991ApJ...376..204M} while exchanging angular momentum with the interior~\citep{2008A&A...489L..53B,2013A&A...556A..36G,2015A&A...577A..98G}.
Another cause could be a change in the nature of emerging photospheric magnetic field and/or the stellar corona~\citep{1984A&A...133..117V,2003A&A...397..147P,2011ApJ...743...48W,2012ApJ...746...43R,2014MNRAS.444.3517M} around \(\Omega_\Sat\).
\citet{1991ApJ...376..204M}, for example, suggested that the surface magnetic field strength \(B\) may saturate in rapid rotators. Alternatively, the knee could be a consequence of a change in the magnetic field complexity~\citep{2015ApJ...807L...6G,2018ApJ...862...90G,2018ApJ...854...78F}.
\citet{2019ApJ...886..120S}, however, found that that field complexity plays a modest role for reasonable mass-loss rates.

In Fig.~\ref{fig:detailed-breakdown} we see that the plotted quantities \(\Omega / \dot \Omega_\Wind\), \(\dot J_\Wind\), \(\dot M_\Wind\) and \(\Phi\) exhibit a similar structure, in which the quantity saturates for \(\Omega \gtrsim \Omega_\Sat\). This suggests that the saturation of the wind torque and wind mass loss rate is a consequence of the saturation of the surface unsigned magnetic flux \(\Phi\) in our models.
The result is consistent with the findings of~\citet{paper3} where no break was found when considering \(\dot J_\Wind\) and \(\dot M\) as a function of surface unsigned magnetic flux or field complexity, and the residual spread was attributed to stellar magnetic cycles.
We thus attribute the saturation in Fig.~\ref{fig:detailed-breakdown} to a saturation in the surface unsigned magnetic flux (and surface magnetic field strength since \(\Phi=4\pi R^2\,B_r\)) for \(\Omega \gtrsim \Omega_\Sat\).
The mass loss rate \(\dot M\) saturation arises naturally in our models, while mass loss rate saturation at high rates of rotation is externally imposed in many models of spin-down~\citep{2013A&A...556A..36G,2015A&A...577A..98G,2015A&A...577A..28J,2017MNRAS.472.2590S,2020A&A...635A.170A}. Such mass loss rate saturation may (or may not) have observational support~\citep{2021LRSP...18....3V,2021ApJ...915...37W}. In the Alfvén wave based model used in this work there is an approximate power-law relationship between \(\dot M\) and \(\Phi\)~\citep{paper3} which means that the mass loss rate saturation is a consequence of magnetic saturation. Similar mass loss rate saturation may be seen also in the one-dimensional Alfvén wave based model of~\citet{2011ApJ...741...54C}, and in the models of~\citet{2011SSRv..158..339S,2013PASJ...65...98S,2020ApJ...896..123S}.

Stellar magnetic activity is known to saturate with increasing rotation~\citep{1984ApJ...279..763N,2003A&A...397..147P} and there are signs of saturation of the surface magnetic field strength~\citep{2009ApJ...692..538R,2014MNRAS.441.2361V}.
This is supported by the saturation in spot filling factors found by \citet{2022MNRAS.517.2165C}.
Early models~\citep{1991ApJ...376..204M} suggested that
that \(B \propto \Omega^a\) where \(a\approx1\) (linear dynamo) in the un-saturated regime and \(\approx0\) (saturated dynamo) in the saturated regime.
In our dataset we find that the surface unsigned magnetic flux in the un-saturated regime \(\Phi\propto \Omega^{1.25\pm0.07}\) is matched by the \(B_r \propto \Omega^{1.32\pm0.14}\) result of \citet{2014MNRAS.441.2361V}. In comparison with the spin-down timescale and mass loss rate values, the surface unsigned magnetic flux appears to continue increasing past \(\Omega_\Sat\), albeit at a slower rate as in~\citet{2014ApJ...794..144R,2022A&A...662A..41R}.

\label{sec:torque-scaling} We have applied a scaling on the wind torque values \(\dot J_\Wind\) in order to match the gyrotrack torque \(\dot J_\Obs\) at solar maximum. In our work the scaling factor has the same value of 7 for all the stars modelled (and the Sun), and is thus independent of \(\Omega\).
Similar torque scalings have a long history in the literature~\citep{1997ApJ...480..303K}.
Solar wind torque estimates based on the solar surface magnetic field strength can be \qty{\sim 20} times smaller than the torque required by the Skumanich relation and the \(\Omega_\Obs(t)\) gyrotracks~\citep{2018ApJ...854...78F,2018ApJ...864..125F}. In \citet{2019ApJ...886..120S} the torque values of~\citet{2018ApJ...854...78F} were scaled by \qty{\sim 25} to match the gyrotracks of~\citet{2015ApJ...799L..23M}. Similarly, the~\citet{2012ApJ...754L..26M} wind torque formula, which was used in the~\citet{2013A&A...556A..36G} gyrotracks and by~\citet{2015A&A...577A..28J} was scaled by a factor of \qty{\sim 2} and \qty{\sim 11} respectively to match cluster rotation rate observations.
In this work, we find that a single, constant \(7\times\) torque scaling factor is sufficient for our Alfvén based wind model to agree with observations.

The need for the \(7\times\) torque scaling may be rooted in underestimates of the stellar magnetic field strengths~\citep{2010MNRAS.407.2269M,2015ApJ...813L..31Y,2019MNRAS.483.5246L} and solar magnetic field strengths, the latter in which a scaling of 2--4 is often applied to match the measured magnetic field at \qty{1}{\astronomicalunit}~\citep{2006ApJ...645.1537C,2013ApJ...778..176O,2017ApJ...848...70L,2018ApJ...856...53P,2018ApJ...864..125F,2019ApJ...887...83S,2021A&A...650A..19R}.
Zeeman-Doppler imaging reconstructs large-scale magnetic features on the stellar surface in an \(\Omega\)-dependent way~\citep{2000MNRAS.318..961H,2010MNRAS.407.2269M}, but only a fraction of the field strength is recovered ~\citep[e.g.\ ][]{2019MNRAS.483.5246L}.
These two biases may be used to justify a scaling of \(\dot J_\Wind\) including the \(7\times\) scaling applied in this work.
On a similar note we find that scaling the surface magnetic field by a factor of \qty{\sim 5} as in \citet{paper1,paper2,paper3} we also obtain gyrotracks with both exponential and Skumanich-type spin-down. We do not pursue this alternative approach here as the resulting wind mass loss rate values differ for the solar case~\citep[see][]{paper3}. We do however emphasise that the qualitative indications of a saturated and an un-saturated magnetic regime also appear in this model series.

\label{sec:core-envelope-decoupling-during-spin-down}
Alternatively, the need for the \(7 \times\) torque scaling may be attributed to core-envelope decoupling~\citep{1986PASP...98.1233S,2010ApJ...716.1269D,2013A&A...556A..36G,2015A&A...577A..98G}, where the stellar convective envelope  may spin at a different rate than the core while exchanging rotational momentum with it at a (configurable) coupling timescale \(\tau\).
Core-envelope decoupling would change the direction of the coloured arrows in Fig.~\ref{fig:trend-omega-age} and, depending on the length of \(\tau\), it could have the same effect as changing the torque scaling factor or scaling the ZDI magnetic field strength.
A variable \(\tau\) was invoked by \citet{2013A&A...556A..36G,2015A&A...577A..98G} to create their observed-based gyrotracks extending from the stellar contraction era to the Sun's age.
In the \citet{2015A&A...577A..42B} models \qty{\sim 20}{\percent} of our stars' rotational inertia is in the envelope. This could produce an effect similar to required torque scaling for long \(\tau\), in that the envelope spins down with little angular momentum exchange with the core in parts of the stars' life.
Core-envelope decoupling has theoretical support from hydrodynamic modelling~\citep{1981ApJ...243..625E,1988AJ.....95.1895R,1989ApJ...338..424P,1992A&A...265..115Z,2016A&A...587A.105A,2019A&A...631A..77A} but it has been suggested  that the inclusion of magnetohydrodynamic instabilities such as the \citet{1973MNRAS.161..365T} instability may effectively flatten the internal rotation curve~\citep{2019MNRAS.485.3661F,2022NatAs...6..788E}. Some models of stellar spin-down~\citep{1994MNRAS.269.1099C,2015ApJ...799L..23M,2015A&A...577A..28J} have applied a model of solid body rotation.
The recently discovered `stalled spin-down'~\citep{2018ApJ...862...33A,2019ApJ...879...49C,2019ApJ...879..100D,2022ApJ...933..114D} of K dwarfs is however attributed to core-envelope decoupling~\citep[e.g.][]{2020A&A...636A..76S,2022ApJ...924...84C,2023ApJ...951L..49C}, although a magnetic wind effect may also play a role~\citep{2020ApJ...904..140C}.

\section{Conclusions}\label{sec:Conclusions}
We have modelled the stellar spin-down rates of 27 young, solar-type stars based on wind models published in~\citet{paper1,paper2,paper3} and whose surface magnetic fields were reconstructed by~\citet{2016MNRAS.457..580F,2018MNRAS.474.4956F} using Zeeman-Doppler imaging. We have compared our resulting wind torques and spin-down rates with the observation-based \(\Omega(t)\) gyrotracks of~\citet{2013A&A...556A..36G}.

Unlike many previous works, our wind torques are derived from fully three-dimensional wind models where the corona is heated by Alfvén waves. As our wind model~\citep{2013ApJ...764...23S,2014ApJ...782...81V} extends into the chromosphere, the coronal density and temperature emerge naturally as a consequence of the coronal heating by Alfvén wave turbulence. This means that we do not need to impose a desired mass loss rate by adjusting the coronal density, as is required in polytropic wind models~\citep{2009ApJ...699..441V,2012ApJ...754L..26M,2015ApJ...798..116R,2017ApJ...845...46F,2018ApJ...864..125F}. Instead, the mass loss rate emerges naturally as a consequence of the coronal heating, and ultimately the surface magnetic field strength.

We find that the torque and the mass loss rate produced by our wind model (using the same solar configuration parameters across the entire age and rotation rate range)
are compatible with features of the observationally derived gyrotracks \(\Omega_\Obs(t)\) for both fast and slow rotators from the onset of the spin-down era until the Sun's age. We find signs of both the exponential (\(\dot \Omega \propto \Omega\)) spin-down rate for rapid rotators and the Skumanich-type (\(\dot \Omega \propto \Omega^3\)) spin-down rate for slow rotators.
In our models the two regimes arise as a consequence of the ZDI observed magnetic field strength saturation at rapid rates of rotation.

Our model is novel as it uses only a single free scaling parameter in order to agree with observations. By scaling all the wind torque based spin-down rates \(\dot \Omega_\Wind\) by a factor of 7, so that the solar maximum value matches the \(\dot \Omega_\Sun\) value predicted from the~\citet{1972ApJ...171..565S} relation \(\Omega\propto t^{-1/2}\), we find good agreement between the gyrotrack-derived spin-down time-scales  \(\Omega / \dot \Omega_\Obs\) and the wind torque-derived spin-down time-scales \(\Omega / \dot \Omega_\Wind\).
Such scalings have long been used in evolutionary models~\citep[e.g.][]{1997ApJ...480..303K}.
The scaling may be needed because of changes in the effective rotational inertia due to core-envelope decoupling, or
due to ZDI underestimates of the unsigned magnetic flux; both effects would lead to an increase in \(\dot \Omega_\Wind\).

Our work suggests that the wind torque saturation is caused by a saturation of the surface magnetic field \emph{strength}, rather than a qualitative change in the magnetic field complexity for rapidly rotating stars.
To further confirm or refute our findings it would be necessary to have more ZDI data of stars with very well constrained ages. ZDI data of stars with rotation periods in the range \qtyrange{0.2}{1}{\day} and \qty{>10}{\day} would be of particular interest.

\section*{Acknowledgements}

    This project has received funding from the European Research Council (ERC) under the European Union's Horizon 2020 research and innovation programme (grant agreement No 817540, ASTROFLOW).
    This work used the Dutch national e-infrastructure with the support of the SURF Cooperative using grant nos. EINF-2218 and EINF-5173.
    This research has made use of NASA's \href{https://ui.adsabs.harvard.edu/}{Astrophysics Data System}.
    This work was carried out using the \swmf{} tools developed at The University of Michigan \href{https://spaceweather.engin.umich.edu/the-center-for-space-environment-modelling-csem/}{Center for Space Environment Modelling (CSEM)} and made available through the NASA \href{https://ccmc.gsfc.nasa.gov/}{Community Coordinated Modelling Center (CCMC)}.
    This work has made use of the following additional numerical software, statistics software and visualisation software:
          NumPy~\citep{2011CSE....13b..22V},
          SciPy~\citep{2020SciPy-NMeth},
     Matplotlib~\citep{2007CSE.....9...90H},
     and
    statsmodels~\citep{seabold2010statsmodels}.
    We would like to express our gratitude to the anonymous reviewers for their valuable contributions, which have enhanced the quality and rigour of our article.

\section*{Data availability}
The data underlying this article will be shared on reasonable request to the corresponding author.

\bibliographystyle{mnras}
\bibliography{bibliography} %

\appendix

\section{Wind models}\label{sec:wind-models}
This work is based upon numerical modelling carried out in \citet{paper1,paper2,paper3} where we used the Zeeman-Doppler imaging observations and surface magnetic maps of~\citet{2016MNRAS.457..580F,2018MNRAS.474.4956F} to drive a three-dimensional numerical wind model incorporating Alfvén wave heating of the corona~\citep[the Alfvén wave solar model \awsom{},][]{2013ApJ...764...23S,2014ApJ...782...81V}. Table~\ref{tab:observed_quantities} gives the relevant stellar parameters used in our modelling, as well as key results from the wind modelling.

The \awsom{} model is part of the block-adaptive tree solarwind Roe upwind scheme \batsrus{}~\citep{1999JCoPh.154..284P,2005JGRA..11012226T} and the space weather modelling framework \swmf{}~\citep{2012JCoPh.231..870T,2021JSWSC..11...42G}\footnote{The open source version of the space weather modelling framework can be found on-line at
\url{https://github.com/MSTEM-QUDA/SWMF.git}. In this work version \href{https://github.com/MSTEM-QUDA/SWMF/commit/a5beb110f}{2.40.{\tt a5beb110f}} (2022-10-08) has been used for all processing.}.
In the \awsom{} model the corona is heated by the dissipation of Alfvén wave energy in the form of an Alfvén wave energy flux emanating from the chromospheric base. The numerical wind models extend from the chromosphere to about \qty{1}{\astronomicalunit}.

At the inner boundary the chromospheric temperature is set to  \(T=\qty{5e4}{\kelvin}\) and the number density is set to \(n=\qty{2e17}{\per\cubic\meter}\)~\citep[following e.g.\ ][]{2018LRSP...15....4G}. These values are used for all our model runs (as described in Table~\ref{tab:parameters}).

The parameters given here are tuned to reproduce solar wind conditions.
In contrast to polytropic wind models, the Alfvén wave based wind model~\citep[][]{2013ApJ...764...23S,2014ApJ...782...81V} effectively shifts the model boundary downwards into the much colder and denser stellar chromosphere, making the coronal density and temperature more dependent on the Alfvén wave based heating model than on the chromospheric boundary conditions. This also permits the consistent computation of the coronal wind density and \(\dot M\) consistently with \(\dot J\) and \(R_\Alfven\) \citep[e.g.\ ][]{2015ApJ...807L...6G,2016A&A...588A..28A,2017ApJ...835..220C}.

The Alfvén wave energy flux across the stellar surface
is proportional to the local magnetic field value~\citep{2014ApJ...782...81V}. The proportionality constant \(\Pi_\Alfven / B\) is set to \qty{1.1e6}{\watt\per\square\meter\per\tesla} which is the standard value used for the Sun and solar-type stars~\citep{2018LRSP...15....4G}.
We note that the Alfvén wave flux applied at the model boundary affects \(\dot M\) and \(\dot J\) differently~\citep{2020A&A...635A.178B,2021ApJ...916...96A,2023ApJ...946L..47H,2023A&A...678A.152V}. By scaling the Alfvén wave flux and the magnetic field strength \(\dot M\) can be varied while \(\dot J\) remains constant.  Recently, the solar Alfvén flux-to-field ratio has been found to be correlated with the area of open field regions~\citep{2023ApJ...946L..47H}. The study also finds the parameter is anti-correlated with the average unsigned radial field in the open regions. Given the relation between \(\dot M\) and \(\Pi_\Alfven/B\) of~\citet{2020A&A...635A.178B} this would reduce the span of \(\dot M\) by a factor of \num{\sim 4} for the solar mass loss shown in Fig.\ref{fig:detailed-breakdown}. Future observations of chromospheric turbulence and solar modelling may justify such a a similar scaling approach for stellar winds.

\subsection{Stellar wind model input data}
The following fundamental stellar parameters are varied between models: the stellar mass \(M\), radius \(R\), and rotation period \(P_\Rot = 2\pi/\Omega\). In our work these values are taken from~\citet{2016MNRAS.457..580F,2018MNRAS.474.4956F} and provided for convenience in Table~\ref{tab:observed_quantities}. The stellar type and age in Table~\ref{tab:observed_quantities} are also from~\citet{2016MNRAS.457..580F,2018MNRAS.474.4956F} but are not in any way part of the wind model input. The solar minimum and solar maximum models use processed GONG magnetograms of Carrington rotation 2211 and 2157 respectively \citep[see][]{paper1}.
The stellar mass, radius, and period of rotation values have uncertainties of
\(\lesssim \qty{10}{\percent}\)  %
\(\lesssim \qty{20}{\percent}\),  %
and
\(\lesssim \qty{10}{\percent}\).  %
The period error is approximately equal to the width of the round symbols used to denote each star in Fig.~\ref{fig:trend-omega-age}, Fig.~\ref{fig:detailed-breakdown} and Fig.~\ref{fig:detailed-breakdown-alfven}.

The surface magnetic maps, in the form of spherical harmonics coefficients of the radial magnetic field, are ingested into the model at the inner boundary. The maps of the radial magnetic components can be seen in~\citet{2016MNRAS.457..580F,2018MNRAS.474.4956F}, in which they were derived, and are also reproduced in \citet{paper1,paper2,paper3}.

\subsection{Stellar wind model output data}
As the \awsom{} numerical model is relaxed towards a steady state, the non-radial magnetic field components are permitted to vary, while the radial magnetic field component stays fixed. Thus the average unsigned surface radial magnetic field \(B_r\) in Table~\ref{tab:observed_quantities} and the unsigned surface magnetic flux \(\Phi=4\pi R^2 B_r\) of Fig.~\ref{fig:detailed-breakdown} (bottom panel) stay fixed for each wind model.

The wind mass loss rate \(\dot M\), wind angular momentum loss rate \(\dot J_\Wind\) and average Alfvén radius \(R_\Alfven\) in Table~\ref{tab:observed_quantities} are computed from the wind model output in~\citet{paper3} and briefly described here for convenience.

We find the wind mass loss rate \(\dot M\) by integrating over a sphere \(S\) centred on the star:
\begin{equation}\label{eq:mass_loss_integral}
    \dot M = \oint\nolimits_S \rho \vec u \cdot \uvec n \,\mathrm{d} S.
\end{equation}
where \(\rho\) and \(\vec u\) are the local wind density and velocity values and \(\uvec n\) is the normal of the surface \(S\).

We compute the angular momentum loss rate \(\dot J_\Wind\) from
\begin{equation}
    \label{eq:angmom_loss}
    \dot J
    = \oint\nolimits_{S}
    \left(\vec B \times \vec r\right)_{3} \left(\frac{\vec B \cdot \uvec r}{\mu_0} \right)
    +
    \left(
        \Omega \varpi^2 + \left(
            \vec r \times \vec v
        \right)_3
    \right)
    \left(
        \vec v \cdot \uvec r
    \right)
    \rho
    \, {\rm d} S.
\end{equation}
Here \(\vec B\) is the local vector magnetic field, \(\mu_0\) is the magnetic constant,
\(\vec \varpi = \vec{r} - (\vec r \cdot \uvec z) \uvec{z}\) denotes the cylindrical distance from the origin, and
\(\vec v = \vec u - \Omega \uvec z \times \vec r\) denotes the wind velocity in a co-rotating frame. \((\cdot)_3\) refers to the \(z\) component vector enclosed. The scalar angular velocity \(\Omega\) is taken from Table~\ref{tab:observed_quantities}.

\begin{figure}
    \centering
    \includegraphics{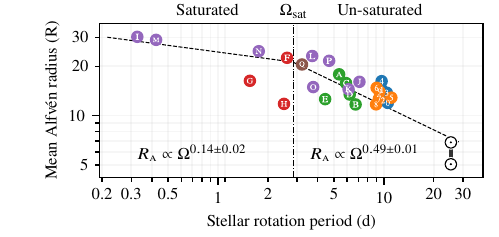}
    \caption{
        Mean Alfvén radius \(R_\Alfven\) plotted against the stellar rotation period as in Fig.~\ref{fig:detailed-breakdown}.
        The fit parameters for \(R_\Alfven\) are given in Table~\ref{tab:fit-coefficients}.
    }\label{fig:detailed-breakdown-alfven}
\end{figure}

We calculate the average Alfvén radius \(R_\Alfven\) by defining the radial distance to the Alfvén surface \(\vec r_\Alfven(\theta, \varphi)\) over each point \(\theta, \varphi\) on the stellar surface. The average Alfvén radius (in stellar radii) is then given by
\begin{equation}
    R_\Alfven = \frac{1}{4\pi R^2}\oint\nolimits_{S} \left|\vec r_\Alfven(\theta, \varphi) \right| \, \mathrm{d}S .
\end{equation}
where \(S\) is the stellar surface and \(R\) is the stellar radius from Table~\ref{tab:observed_quantities}. \(R_\Alfven\) is thus the spherically averaged value of the distance to the Alfvén surface over the spherical surface parametrised by the polar and azimuthal coordinates \((\theta, \varphi)\). Fig.~\ref{fig:detailed-breakdown-alfven} shows the average Alfvén radius \(R_\Alfven\) plotted against the stellar rotation period as in Fig.~\ref{fig:detailed-breakdown}.

\section{Spin-down timescales}\label{sec:spin-down-timescales}
To compute the spin-down timescales \(\Omega / \dot \Omega_\Wind\) we use the scaled wind torque \(\dot J_\Wind\) values from \citet{paper3} and the rotational inertia \(I\) from~\citet{2015A&A...577A..42B}. In this model the solar rotational inertia \(I_\Sun=\qty{7.1e46}{\kilo\gram\metre\squared}\). The exact values used for each star is given in Table~\ref{tab:observed_quantities}.  We note that there is some spread in reported solar rotational inertia values, as \(I_\Sun=\qty{5.8e46}{\kilogram\meter\squared}\) in e.g.\ \citet{2000asqu.book.....C}.

The rotational inertia \(I\) values are not part of the wind model input, but are used to compute the spin-down timescales \(\Omega / \dot \Omega_\Wind\) as described in Section~\ref{sec:wind-torques} with equation~\eqref{eq:omegadot-from-torque}. Note that we do not consider the contraction torque term \(\Omega \dot I\) in this work so that in effect \(\dot J_\Wind = \dot \Omega_\Wind I\) in equation~\eqref{eq:omegadot-from-torque}. In this work we also do not consider core-envelope decoupling \citep[e.g.][]{1986PASP...98.1233S}.
The spin-down timescales \(\Omega / \dot \Omega_\Wind\)
 for each star are computed from the values given in Table~\ref{tab:observed_quantities}. These values are also plotted in Fig.~\ref{fig:trend-omega-age} and the top panel of Fig.~\ref{fig:detailed-breakdown}.

\begin{figure}
    \includegraphics{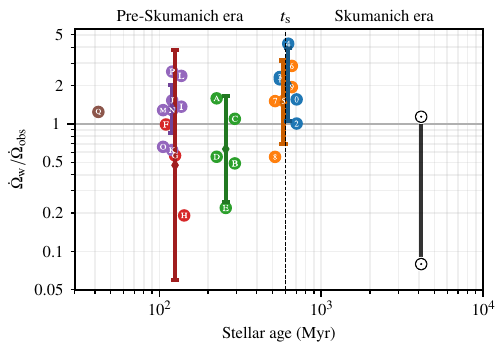}
    \caption{Residual ratio \(\dot \Omega_\Wind / \dot \Omega_\Obs \).
    The  stellar symbol represent solar maximum and minimum.
    Values smaller than unity are associated with reduced rotational braking in Fig.~\ref{fig:trend-omega-age}.
    The dark vertical lines show \qty{95}{\percent} confidence intervals for the average  \(\dot \Omega_\Wind / \dot \Omega_\Obs \) in each cluster. Numerical values pertaining to this plot is given in Table~\ref{tab:results-cluster}.
    }\label{fig:trend-omega-age-residuals}
\end{figure}

 Fig.~\ref{fig:trend-omega-age-residuals} shows the residual ratio \(\dot \Omega_\Wind / \dot \Omega_\Obs \) for each star. The residual ratio is computed from the gyrotrack spin-down timescale \(\Omega / \dot \Omega_\Obs\) and the wind spin-down timescale \(\Omega / \dot \Omega_\Wind\). The numerical values of \(\Omega / \dot \Omega_\Obs\), \(\Omega / \dot \Omega_\Wind\), and \(\dot \Omega_\Wind / \dot \Omega_\Obs \) are given in Table~\ref{tab:results}.
The dark, coloured bars in Figure \ref{fig:trend-omega-age-residuals} are \qty{95}{\percent} confidence intervals of the mean \(\dot \Omega_\Wind / \dot \Omega_\Obs \) in each cluster. The confidence intervals are computed in the standard way~\citep[e.g.][]{1998ara..book.....D} using the Student's \(t\) distribution with \(n-1\) degrees of freedom, where \(n\) is the number of stars in each cluster.
For the solar minimum and maximum models we do not compute or show a confidence interval as there is only one model for each of these cases. Similarly for \C1Q no confidence interval can be computed as we only have one model aged \qty{42\pm6}{\mega\year}. The confidence interval parameters are given in Table~\ref{tab:results-cluster}.

The confidence intervals can be seen to cover the \(\dot \Omega_\Wind / \dot \Omega_\Obs = 1\) line in Fig.~\ref{fig:trend-omega-age-residuals} for each of the clusters except the Hyades where the confidence interval ends just above the line. Tentatively, it appears that the Hercules-Lyra cluster (green symbols) has a lower \(\dot \Omega_\Wind / \dot \Omega_\Obs\) ratio than the other clusters. However, the confidence intervals are wide and the difference is not statistically significant. Similarly a trend of increasing \(\dot \Omega_\Wind / \dot \Omega_\Obs\) with age is not statistically significant. We attribute the residual spread in \(\dot \Omega_\Wind / \dot \Omega_\Obs\) to stellar magnetic cycles~\citep[see][]{paper1} affecting the surface magnetic flux \(\Phi \) (a source of scatter in the bottom panel of Fig.~\ref{fig:detailed-breakdown}), but systematic errors in magnetogram detail with increasing \(\Omega\)~\citep{2010MNRAS.407.2269M} may also play a role, as may stellar mass differences (see~Table~\ref{tab:observed_quantities}).
\begin{table}
    \centering
    \sisetup{
        table-figures-decimal=2,
        table-figures-integer=2,
        table-figures-uncertainty=2,
        table-figures-exponent = 0,
        table-number-alignment=center,
        round-mode=places,
        round-precision=1
    }
    \caption{
        Spin-down timescales computed in this work, and the residual ratio between the gyro-track spin-down timescale and the wind spin-down timescale. The residual ratios are plotted against stellar age in in Fig.~\ref{fig:trend-omega-age-residuals}.
    }\label{tab:results}
    \begin{tabular}{
    l
    S[table-format=1.2e2, round-precision=2]
    S[table-format=1.2e2, round-precision=2]
    S[table-format=1.2, round-precision=2]
}

\toprule
Case
& {\(\Omega / \dot\Omega_\text{obs}\)}
& {\(\Omega / \dot\Omega_\textsc{w}\)}
& {\(\dot\Omega_\textsc{w} / \dot\Omega_\text{obs}\)}
\\ %
& {\( \left(  \si{\year}                 \right) \)}
& {\( \left(  \si{\year}                 \right) \)}
&
\\ 
\midrule
\SimSymbol{1}{0} Mel25-5 & 1.60e+09   & 1.02e+09   & 1.558     \\
\SimSymbol{1}{1} Mel25-21 & 1.35e+09   & 6.04e+08   & 2.239     \\
\SimSymbol{1}{2} Mel25-43 & 1.40e+09   & 1.39e+09   & 1.008     \\
\SimSymbol{1}{3} Mel25-151 & 1.55e+09   & 6.63e+08   & 2.336     \\
\SimSymbol{1}{4} Mel25-179 & 1.34e+09   & 3.15e+08   & 4.263     \\
\midrule
\SimSymbol{1}{5} AV 523 & 1.76e+09   & 1.16e+09   & 1.518     \\
\SimSymbol{1}{6} AV 1693 & 1.17e+09   & 4.10e+08   & 2.854     \\
\SimSymbol{1}{7} AV 1826 & 1.25e+09   & 8.24e+08   & 1.512     \\
\SimSymbol{1}{8} AV 2177 & 1.15e+09   & 2.09e+09   & 0.552     \\
\SimSymbol{1}{9} TYC 1987 & 1.27e+09   & 6.53e+08   & 1.946     \\
\midrule
\SimSymbol{1}{A} DX Leo & 4.13e+08   & 2.60e+08   & 1.589     \\
\SimSymbol{1}{B} EP Eri & 6.53e+08   & 1.33e+09   & 0.491     \\
\SimSymbol{1}{C} HH Leo & 5.00e+08   & 4.55e+08   & 1.097     \\
\SimSymbol{1}{D} V439 And & 5.54e+08   & 1.00e+09   & 0.553     \\
\SimSymbol{1}{E} V447 Lac & 2.80e+08   & 1.27e+09   & 0.220     \\
\midrule
\SimSymbol{1}{F} HII 296 & 1.15e+08   & 1.17e+08   & 0.991     \\
\SimSymbol{1}{G} HII 739 & 1.15e+08   & 2.04e+08   & 0.565     \\
\SimSymbol{1}{H} PELS 031 & 1.15e+08   & 6.02e+08   & 0.192     \\
\midrule
\SimSymbol{1}{I} BD-072388 & 1.15e+08   & 8.45e+07   & 1.367     \\
\SimSymbol{1}{J} HD 6569 & 7.26e+08   & 4.75e+08   & 1.530     \\
\SimSymbol{1}{K} HIP 10272 & 5.37e+08   & 8.60e+08   & 0.624     \\
\SimSymbol{1}{L} HIP 76768 & 1.96e+08   & 8.20e+07   & 2.385     \\
\SimSymbol{1}{M} LO Peg & 1.15e+08   & 8.99e+07   & 1.284     \\
\SimSymbol{1}{N} PW And & 1.15e+08   & 9.00e+07   & 1.282     \\
\SimSymbol{1}{O} TYC 0486 & 2.01e+08   & 3.03e+08   & 0.662     \\
\SimSymbol{1}{P} TYC 5164 & 3.13e+08   & 1.21e+08   & 2.575     \\
\midrule
\SimSymbol{1}{Q} BD-16351 & 1.47e+08   & 1.18e+08   & 1.244     \\
\midrule
{\(\odot\)} Sun CR2157 & 9.20e+09   & 8.08e+09   & 1.139     \\
{\(\odot\)} Sun CR2211 & 9.20e+09   & 1.16e+11   & 0.080     \\
\bottomrule
\end{tabular}

\end{table}
\begin{table}
    \centering
    \caption{
        Residual ratio statistics for each cluster. For each cluster of this study we give the age, followed by the (dimensionless) mean residual ratio \(\dot \Omega_\Wind / \dot \Omega_\Obs \) and the \qty{95}{\percent} confidence interval of the mean. These parameters can also be seen visualised in Fig.~\ref{fig:trend-omega-age-residuals}. In the solar case the range between our minimum and maximum models is \(0.080 < \dot \Omega_\Wind / \dot \Omega_\Obs < 1.14\).
    }\label{tab:results-cluster}
    \sisetup{
        table-figures-decimal=2,
        table-figures-integer=2,
        table-figures-uncertainty=2,
        table-figures-exponent = 0,
        table-number-alignment=center,
        round-mode=places,
        round-precision=1
    }
    \begin{tabular}{
        l
        S[table-format=3(2), round-precision=0]
        S[table-format=1.2, round-precision=2]
        S[table-format=1.2, round-precision=2]
        S[table-format=1.2, round-precision=2]
    }
        \toprule
        Cluster &
        {Age} &
        \multicolumn{3}{c}{Residual ratio \(\dot \Omega_\Wind / \dot \Omega_\Obs \)}\\
        & {(Myr)} &  {2.5\%} & {Mean}  & {97.5\%} \\
        \midrule
        \Unscaled{~}{C5}\,Col-Hor-Tuc    &  42 \pm  6  &{---} &  1.24 & {---} \\
        \Unscaled{~}{C4}\,AB Doradus     & 120 \pm 10  &0.85 &  1.31 & 2.02 \\
        \Unscaled{~}{C3}\,Pleiades       & 125 \pm  8  &0.06 &  0.48 & 3.78 \\
        \Unscaled{~}{C2}\,Hercules-Lyra  & 257 \pm 46  &0.25 &  0.64 & 1.65 \\
        \Unscaled{~}{C1}\,Coma Berenices & 584 \pm 10  &0.69 &  1.48 & 3.14 \\
        \Unscaled{~}{C0}\,Hyades         & 625 \pm 50  &1.05 &  2.04 & 3.95 \\
        \bottomrule
    \end{tabular}
\end{table}

\bsp	%
\label{lastpage} %

\clearpage
\end{document}